\documentclass[prd, nofootinbib, aps, twocolumn, 
showpacs, preprintnumbers, amsmath, amssymb, floatfix]{revtex4-1}
\usepackage{graphicx}
\usepackage{booktabs}
\usepackage{bbm}
\newcommand{\eV}{\mathrm{eV}}
\newcommand{\meV}{\mathrm{meV}}

\newcommand{\GeV}{\mathrm{GeV}}

\newcommand{\kcut}{\ensuremath{k_{\text{cut}}}}
\newcommand{\axion}{\ensuremath{a}}
\newcommand{\electron}{\ensuremath{e^-}}
\newcommand{\ax}{\ensuremath{_{a}}}
\newcommand{\fax}{\ensuremath{f_{\mathrm{PQ}}}}

\newcommand{\maxion}{\ensuremath{m_{a}}}

\newcommand{\be}{\begin{equation}}
\newcommand{\ee}{\end{equation}}
\newcommand{\benn}{\begin{equation*}}
\newcommand{\eenn}{\end{equation*}}
\newcommand{\bea}{\begin{equation}\begin{aligned}}
\newcommand{\eea}{\end{aligned}\end{equation}}
\newcommand{\beann}{\begin{equation*}\begin{aligned}}
\newcommand{\eeann}{\end{aligned}\end{equation*}}
\newcommand{\bse}{\begin{subequations}}
\newcommand{\ese}{\end{subequations}}

\newcommand{\TD}{T_{\mathrm{D}}}
\newcommand{\Reheating}{\mathrm{R}}
\newcommand{\TR}{T_{\Reheating}}
\newcommand{\fb}{\ensuremath{f_{\text{B}}}}

\newcommand{\YaxTP}{Y_{\axion}^{\mathrm{TP}}}

\newcommand{\nax}{n_{\axion}}
\newcommand{\Yaxeq}{Y_{\axion}^{\mathrm{eq}}}
\newcommand{\Omegaaxeq}{\Omega_{\axion}^{\mathrm{eq}}}
\newcommand{\naxeq}{n_{\axion}^{\mathrm{eq}}}
\begin{document}
% __________________________________________________________________
%
% ___ Preprint Numbers ___________________________________________________
%
\preprint{arXiv:1008.4528}
\preprint{MPP-2010-20}
%
%
% ___ Preamble ______________________________________________________
%
% __________________________________________________________________
\title{Thermal axion production in the primordial quark-gluon plasma}
% __________________________________________________________________
\author{Peter Graf}
%\email{graf@mpp.mpg.de}
\affiliation{Max-Planck-Institut f\"ur Physik, 
F\"ohringer Ring 6,
D--80805 Munich, Germany}
% __________________________________________________________________
\author{Frank Daniel Steffen}
%\email{steffen@mpp.mpg.de}
\affiliation{Max-Planck-Institut f\"ur Physik, 
F\"ohringer Ring 6,
D--80805 Munich, Germany}
% __________________________________________________________________
%
% ___ Abstract _________________________________________________________
%
\begin{abstract}
  We calculate the rate for thermal production of axions via
  scattering of quarks and gluons in the primordial quark-gluon
  plasma.
  To obtain a finite result in a gauge-invariant way that is
  consistent to leading order in the strong gauge coupling, we use
  systematic field theoretical methods such as hard thermal loop
  resummation and the Braaten--Yuan prescription.
  The thermally produced yield, the decoupling temperature, and the
  density parameter are computed for axions with a mass below 10~meV.
  In this regime, with a Peccei--Quinn scale above $6\times
  10^8\,\GeV$, the associated axion population can still be
  relativistic today and can coexist with the axion cold dark matter
  condensate.
\end{abstract}
%
%% 04.65.+e     Supergravity (see also 12.60.Jv Supersymmetric models)
%% 12.60.Jv     Supersymmetric models (see also 04.65.+e Supergravity)
%% 14.80.Ly     Supersymmetric partners of known particles
%% 14.80.Va     Axions
%% 95.30.Cq     Elementary particle processes
%% 95.35.+d     Dark matter
%% 98.80.Cq     Particle-theory and field-theory models of the early Universe
%
%\keywords{}
\maketitle

\section{Introduction}

If the Peccei--Quinn (PQ) mechanism is the
explanation of the strong CP problem, axions will pervade the Universe
as an extremely weakly interacting light particle species.
In fact, an axion condensate is still one of the most compelling
explanations of the cold dark matter in our
Universe~\cite{Sikivie:2006ni,Kim:2008hd}.
While such a condensate would form at temperatures $T\lesssim 1~\GeV$,
additional populations of axions can originate from processes at much
higher temperatures.
Even if the reheating temperature $\TR$ after inflation is such that
axions were never in thermal equilibrium with the primordial plasma,
they can be produced efficiently via scattering of quarks and gluons.
Here we calculate for the first time the associated thermal production
rate consistent to leading order in the strong gauge coupling $g_s$ in
a gauge-invariant way.
The result allows us to compute the associated relic abundance and to
estimate the critical $\TR$ value below which our considerations are
relevant. 
For a higher value of $\TR$, one will face the case in which axions
were in thermal equilibrium with the primordial plasma before
decoupling at the temperature $\TD$ as a hot thermal
relic~\cite{Masso:2002np,Sikivie:2006ni}.
The obtained critical $\TR$ value can then be identified with the
axion decoupling temperature~$\TD$.

We focus on the model-independent axion ($a$) interactions with gluons
given by the Lagrangian%
\footnote{The relation of $\fax$ to the vacuum expectation value
 $\langle\phi\rangle$ that
  breaks the U(1)$_{\mathrm{PQ}}$ symmetry depends on the axion model
  and the associated domain wall number $N$:
  $\fax\propto\langle\phi\rangle/N$;
  cf.~\cite{Sikivie:2006ni,Kim:2008hd} and references therein.}
\be
\mathcal{L}\ax 
= 
\frac{g_s^2}{32\pi^2\fax}\, 
a\,G_{\mu\nu}^b\widetilde{G}^{b\,\mu\nu},
\
\label{Eq:AxionInteraction}
\ee
with the gluon field strength tensor $G^b_{\mu\nu}$, its dual
$\widetilde{G}^b_{\mu\nu}=\epsilon_{\mu\nu\rho\sigma}G^{b\,\rho\sigma}/2$,
and the scale $\fax$ at which the PQ symmetry is broken spontaneously.
Numerous laboratory, astrophysical, and cosmological studies point to
\be
\fax \gtrsim 6 \times 10^8~\GeV
\ ,
\label{Eq:fa_Limit}
\ee
which implies that axions are stable on cosmological
timescales~\cite{Raffelt:2006cw,Amsler:2008zzb}.
Considering this $\fax$ range, we can neglect axion production via
$\pi\pi\to\pi\axion$ in the primordial hot hadronic
gas~\cite{Chang:1993gm,Hannestad:2005df}.
Moreover, Primakoff processes such as
$\electron\gamma\to\electron\axion$
or 
$q\,\gamma\rightarrow q\,\axion$
are not taken into account. These processes depend on the axion model
and involve the electromagnetic coupling instead of the strong one
such that they are usually far less efficient in the early
Universe~\cite{Turner:1986tb}.

We assume a standard thermal history and refer to $\TR$ as the initial
temperature of the radiation-dominated epoch.
While inflation models can point to $\TR$ well above $10^{10}~\GeV$,
we focus on the case $\TR<\fax$ such that no PQ symmetry restoration
takes place after inflation.

Related studies exist. The decoupling of axions out of thermal
equilibrium with the primordial quark-gluon plasma (QGP) was
considered in Refs.~\cite{Masso:2002np,Sikivie:2006ni}. While the same
QCD processes are relevant, our study treats the thermal production of
axions that were never in thermal equilibrium. Moreover, we use hard
thermal loop (HTL) resummation~\cite{Braaten:1989mz} and the
Braaten--Yuan prescription~\cite{Braaten:1991dd}, which allow for a
systematic gauge-invariant treatment of screening effects in the QGP.
In fact, that prescription was introduced on the example of axion
production in a hot QED plasma~\cite{Braaten:1991dd}; see
also~Ref.~\cite{Bolz:2000fu}.

\section{Thermal production rate}

\begin{figure}[t]
\begin{description}
\item [Process A] $g^a+ g^b \rightarrow g^c + a$
\begin{center}
\includegraphics[width=0.45\textwidth]{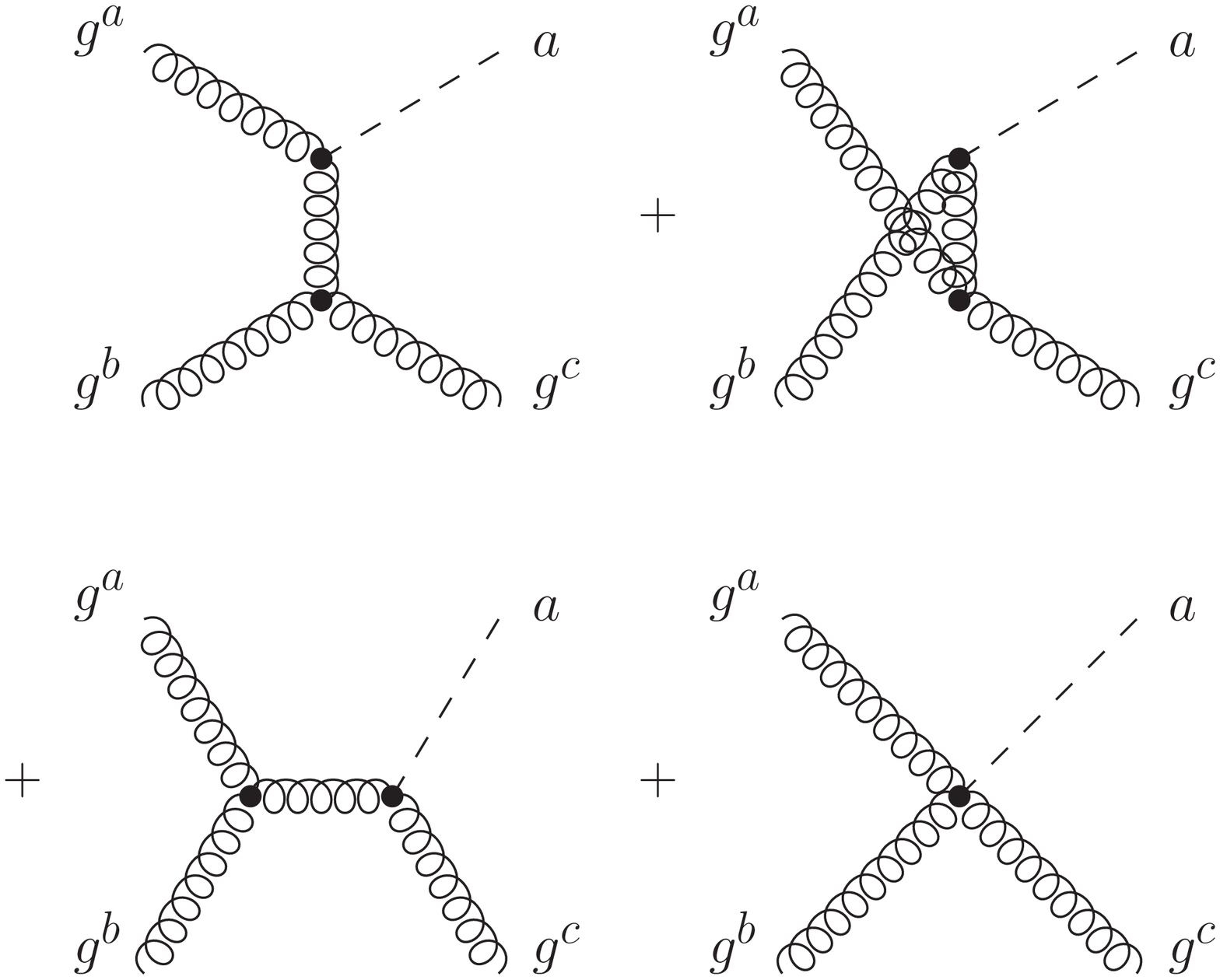}
\end{center}
\item [Process B] $q_i + \bar{q}_j \rightarrow g^a + a$
\begin{center}
\includegraphics[width=0.205\textwidth]{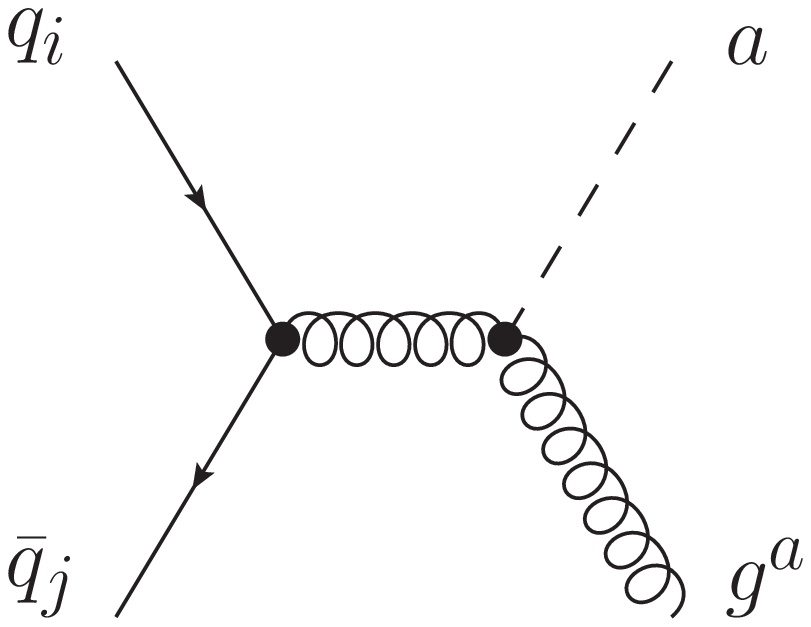}
\end{center}
\item [Process C] $q_i + g^a \rightarrow q_j + a$ (crossing of B)
\end{description}
\caption{The $2\to 2$ processes for axion production in the QGP.
  Process C also exists with antiquarks $\bar{q}_{i,j}$ replacing
  $q_{i,j}$.}
\label{Fig:AxionProduction}
\end{figure}

Let us calculate the thermal
production rate of axions with energies $E\gtrsim T$ in the hot QGP.
The relevant $2\to 2$ scattering processes
involving~(\ref{Eq:AxionInteraction}) are shown in
Fig.~\ref{Fig:AxionProduction}.
The corresponding squared matrix elements are listed in
Table~\ref{Tab:Misquared}, where $s=(P_1 + P_2)^2$ and $t=(P_1 -
P_3)^2$ with $P_1$, $P_2$, $P_3$, and $P$ referring to the particles
in the given order. Working in the limit $T \gg m_i$, the masses of
all particles involved have been neglected.  Sums over initial and
final spins have been performed. For quarks, the contribution of a
single chirality is given.
The results obtained for processes A and C point to potential 
IR divergences associated with the exchange of soft (massless)
gluons in the $t$ channel and $u$ channel. Here screening effects of
the plasma become relevant.
To account for such effects, the QCD Debye mass
$m_{\mathrm D}= \sqrt{3} m_g$,
with 
$
m_g = g_s T \sqrt{N_c + (n_f/2)}/3
$
for $N_c=3$ colors and $n_f=6$ flavors,
was used in Ref.~\cite{Masso:2002np}.
In contrast, our calculation relies on HTL
resummation~\cite{Braaten:1989mz,Braaten:1991dd} which treats
screening effects more systematically.
%
% __________________________________________________________________
\begin{table}[b]
\centering
\caption{Squared matrix elements for axion ($\axion$) production in 2-body processes involving quarks of a single chirality ($q_i$) and gluons ($g^a$) in the high-temperature limit, $T \gg m_i$, with the SU($N_c$) color matrices $f^{abc}$ and $T^a_{ji}$. Sums over initial and final state spins have been performed.}
\label{Tab:Misquared}
\begin{ruledtabular}
\begin{tabular}{@{\extracolsep{\fill}}ccc}
Label $i$ 
& Process $i$ 
& $|M_i|^2 / \left(\frac{g_s^6}{128\pi^4 \fax^2}\right)$ \\ 
\noalign{\smallskip}
\hline
\noalign{\smallskip}
A 
& $g^a + g^b \rightarrow g^c + \axion$ 
& $-4\frac{(s^2+st+t^2)^2}{st(s+t)}|f^{abc}|^2$ \\
B 
& $q_i + \bar{q}_j \rightarrow g^a + \axion$ 
& $\left(\frac{2t^2}{s} + 2t+s\right)|T_{ji}^a|^2$ \\
C 
& $q_i + g^a \rightarrow q_j + \axion$
& $\left(-\frac{2s^2}{t}-2s-t\right)|T_{ji}^a|^2$ \\
\end{tabular}
\end{ruledtabular}
\end{table}
% __________________________________________________________________
%
% __________________________________________________________________
% __________________________________________________________________
%
%

Following Ref.~\cite{Braaten:1991dd}, we introduce a momentum scale
$\kcut$ such that $g_sT\ll\kcut\ll T$ in the weak coupling limit
$g_s\ll 1$. This separates soft gluons with momentum transfer of order
$g_sT$ from hard gluons with momentum transfer of order $T$. By
summing the respective soft and hard contributions, the finite rate
for thermal production of axions with $E\gtrsim T$ is obtained in
leading order in $g_s$, 
\begin{align}
E\frac{dW\ax}{d^3p} 
=
E\frac{dW\ax}{d^3p}\biggr\vert_{\text{soft}} 
+
E\frac{dW\ax}{d^3p}\biggr\vert_{\text{hard}},
\label{Eq:AxionTPRate}
\end{align}
which is independent of $\kcut$; cf.~(\ref{Eq:SoftPart})
and~(\ref{Eq:HardPartResult}) given below.

\begin{figure}[b]
\centering
\includegraphics[width=0.38\textwidth]{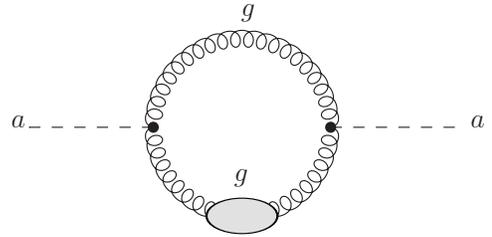}
\caption{Leading contribution to the axion self-energy for soft gluon
  momentum transfer and hard axion energy. The blob on the gluon line
  denotes the HTL-resummed gluon propagator.}
\label{Fig:AxionSelfEnergy}
\end{figure}

In the region with $k<\kcut$, we obtain the soft contribution from the
imaginary part of the thermal axion self-energy with the ultraviolet
cutoff $\kcut$,
\begin{align}
&E\frac{dW\ax}{d^3p}\biggr\vert_{\text{soft}} 
= 
-\frac{\fb(E)}{(2\pi)^3}
\operatorname{Im}\Pi_{\axion}(E+i\epsilon,\vec{p})\vert_{k<\kcut}
\label{Eq:OpticalTheorem}
\\
&=
E\fb(E)\frac{3m_g^2g_s^4(N_c^2-1)T}{8192\pi^8 \fax^2}\!\!
\left[ \ln\!\!\left( \frac{\kcut^2}{m_g^2} \right) -1.379 \right]
\label{Eq:SoftPart}
\end{align}
with the equilibrium phase space density for bosons (fermions)
$f_{\mathrm{B(F)}}(E)=[\exp(E/T)\mp 1]^{-1}$.
Our derivation of~(\ref{Eq:SoftPart}) follows
Ref.~\cite{Braaten:1991dd}. The leading order contribution to
$\operatorname{Im}\Pi_{\axion}$ for $k<\kcut$ and $E\gtrsim T$ comes
from the Feynman diagram shown in Fig.~\ref{Fig:AxionSelfEnergy}.
Because of $E\gtrsim T$, only one of the two gluons can have a soft
momentum. Thus only one effective HTL-resummed gluon propagator is
needed.
%
% __________________________________________________________________
% __________________________________________________________________

In the region with $k>\kcut$, bare gluon propagators can be used since
$\kcut$ provides an IR cutoff. From the results given in
Table~\ref{Tab:Misquared} weighted with appropriate multiplicities,
statistical factors, and phase space densities, we then obtain the
(angle-averaged) hard contribution
\begin{align}
& 
E\frac{dW\ax}{d^3p}\biggr\vert_{\text{hard}}
=
\frac{1}{2(2\pi)^3}
\int\frac{d\Omega_p}{4\pi} 
\int\left[ \prod_{j=1}^3\frac{d^3p_j}{(2\pi)^3 2E_j} \right]  
\nonumber\\
&\times(2\pi)^4 \delta^4(P_1+P_2-P_3-P)\Theta(k-\kcut)
\nonumber\\
&\times \sum f_1(E_1)f_2(E_2)[1\pm f_3(E_3)] | M_{1+2\rightarrow 3+\text{a}} | ^2
\label{Eq:HardPartStart}
\\
&
=E\frac{g_s^6(N_c^2-1)}{512\pi^7 \fax^2} 
\left\{
n_f\frac{\fb(E)T^3}{48\pi}\ln(2)
\right. 
\nonumber\\
&
+\left(N_c+\frac{n_f}{2}\right)\frac{\fb(E)T^3}{48\pi}
\left[ \ln \left(\frac{T^2}{\kcut^2}\right) + \frac{17}{3} - 2\gamma + \frac{2\zeta'(2)}{\zeta(2)} \right] 
\nonumber\\
&
\left. 
+N_c(I_{\text{BBB}}^{(1)}-I_{\text{BBB}}^{(3)})
+n_f(I_{\text{FBF}}^{(1)}+I_{\text{FFB}}^{(3)}) 
\right\}
\label{Eq:HardPartResult}
\end{align}
with Euler's constant $\gamma$, Riemann's zeta
function $\zeta(z)$,
\begin{align}
&
I_{\text{BBB(FBF)}}^{(1)}
= 
\frac{1}{32\pi^3}
\!\int_0^\infty \! dE_3 
\!\int_0^{E+E_3} \! dE_1
\ln\left(\frac{|E_1-E_3|}{E_3}\right) 
\nonumber\\
&  
\times
\left\{- \Theta(E_1-E_3)
\frac{d}{dE_1}\!\!
\left[f_{\text{BBB(FBF)}}\frac{E_2^2}{E^2}(E_1^2+E_3^2)\right] 
\right. 
\nonumber\\
& 
\quad
+ \Theta(E_3-E_1)
\frac{d}{dE_1}
[f_{\text{BBB(FBF)}} (E_1^2+E_3^2)] 
\nonumber\\
& 
\left. 
\quad
+ \Theta(E-E_1)
\frac{d}{dE_1}\!\!
\left[f_{\text{BBB(FBF)}}\!\!\left(\frac{E_1^2E_2^2}{E^2}-E_3^2\right)\right] 
\right\},
\label{Eq:I1}
\end{align}
\begin{align}
&
I_{\text{BBB(FFB)}}^{(3)} 
= 
\frac{1}{32\pi^3} 
\int_0^\infty dE_3 
\int_0^{E+E_3} dE_2 \, f_{\text{BBB(FFB)}} 
\nonumber \\
&
\times 
\left\{ 
\Theta(E-E_3) \frac{E_1^2E_3^2}{E^2(E_3+E)} 
+ 
\Theta(E_3-E) \frac{E_2^2}{E_3+E}
\right. 
\nonumber\\
& 
\quad\,
+ \! [\Theta(E_3-E)\Theta(E_2-E_3)-\Theta(E-E_3)\Theta(E_3-E_2)]
\nonumber\\
& 
\quad\quad\quad
\times \frac{E_2-E_3}{E^2}\,[E_2(E_3-E)-E_3(E_3+E)] 
\biggr\},
\label{Eq:I3}
\\
& 
f_{\text{BBB,FBF,FFB}}
=
f_1(E_1)f_2(E_2)[1\pm f_3(E_3)] 
.
\label{Eq:fXXX}
\end{align}
The sum in~(\ref{Eq:HardPartStart}) is over all axion production
processes $1+2\rightarrow 3+\text{a}$ viable
with~(\ref{Eq:AxionInteraction}). The colored particles 1--3 were in
thermal equilibrium at the relevant times. Performing the calculation
in the rest frame of the plasma, $f_{i}$ are thus described by
$f_{\mathrm{F/B}}$ depending on the respective spins. Shorthand
notation~(\ref{Eq:fXXX}) indicates the corresponding combinations,
where $+$ ($-$) accounts for Bose enhancement (Pauli blocking) when
particle~$3$ is a boson (fermion). 
With any initial axion population diluted away by inflation and $T$
well below $\TD$ so that axions are not in thermal equilibrium, the
axion phase space density $f_{\axion}$ is negligible in comparison to
$f_{\mathrm{F/B}}$. Thereby, axion disappearance reactions ($\propto
f_{\axion}$) are neglected as well as the respective Bose enhancement
($1+f_{\axion} \approx1$).
Details on the methods applied to obtain our
results~(\ref{Eq:HardPartResult})--(\ref{Eq:I3})
can be found in Refs.~\cite{Bolz:2000fu,Pradler:2007ne,Graf:2009da}.

\section{Relic axion abundance}

\begin{figure}[b]
\centering
\includegraphics[width=0.48\textwidth,clip]{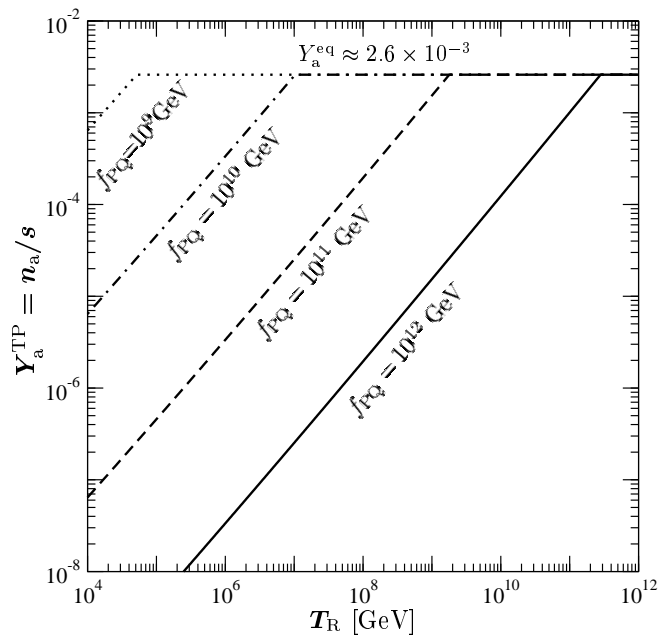}
\caption{The relic axion yield today originating from thermal
  processes in the primordial plasma for cosmological scenarios
  characterized by different $\TR$ values covering the range from
  $10^4$ to $10^{12}\,\GeV$. The dotted, dash-dotted, dashed, and
  solid lines are obtained for $\fax=10^9$, $10^{10}$, $10^{11}$, and
  $10^{12}\,\GeV$.}
\label{Fig:AxionYield}
\end{figure}

We now calculate the thermally produced
(TP) axion yield $\YaxTP=\nax/s$, where $\nax$ is the corresponding
axion number density and $s$ the entropy density.  
For $T$ sufficiently below $\TD$, the evolution of the thermally
produced $\nax$ with cosmic time $t$ is governed by the Boltzmann
equation
\be 
\frac{d\nax}{dt} + 3H\nax 
= 
\int d^3p \,\frac{dW\ax}{d^3p}
=
W_{\axion} .
\label{Eq:BoltzmannEquation}
\ee
Here $H$ is the Hubble expansion rate, and the collision term is the
integrated thermal production rate
\be
W_{\axion} 
= 
\frac{\zeta(3) g_s^6 T^6}{64\pi^7\fax^2} 
\left[ \ln\left(\frac{T^2}{m_g^2} \right) + 0.406 \right] .
\label{collisionSM}
\ee
Assuming conservation of entropy per comoving volume element,
$(\ref{Eq:BoltzmannEquation})$ can be written as
$d\YaxTP/dt=W_{\axion}/s$. Since thermal axion production proceeds
basically during the hot radiation-dominated epoch, i.e., well above
the temperature of radiation-matter equality $T_{\mathrm{mat=rad}}$,
one can change variables from cosmic time $t$ to temperature $T$
accordingly.  With initial temperature $\TR$ at which $\YaxTP(\TR)=0$,
the relic axion yield today is given by
\begin{align}
&\YaxTP
\approx
\YaxTP(T_{\mathrm{mat=rad}})
= \int_{T_{\mathrm{mat=rad}}}^{\TR} dT
\frac{W_{\axion}(T)}{T s(T) H(T)}
\nonumber\\
&= 
18.6 g_s^6 
\ln \! \left(\frac{1.501}{g_s}\right) \!
\left( \frac{10^{10}\,\GeV}{\fax} \right)^{\!\!2} \!
\left( \frac{\TR}{10^{10}\,\GeV} \right)\! .
\label{Eq:AxionYieldTP}
\end{align}
This result is shown by the diagonal lines in
Fig.~\ref{Fig:AxionYield} for cosmological scenarios with different
$\TR$ values ranging from $10^4$ to $10^{12}\,\GeV$.
%
% __________________________________________________________________
% __________________________________________________________________
%
Here we use $g_s=g_s(\TR)$ as described by the 1-loop renormalization
group evolution~\cite{Amsler:2008zzb}
\be
g_s(\TR)
=
\!\!
\left[
g_s^{-2}(M_Z) 
+ 
\frac{11 N_c-2 n_f}{24\pi^2}
\ln\!\left(\frac{\TR}{M_Z}\right)
\right]^{\!-1/2}
\label{Eq:gS_SM}
\ee
where $g_s^{2}(M_Z)/(4\pi)=0.1172$ at $M_Z=91.188~\GeV$.
Since the methods~\cite{Braaten:1989mz,Braaten:1991dd} allowing for
the gauge-invariant derivation require $g_s\ll 1$,
\eqref{Eq:AxionYieldTP} is most reliable for
$\TR\gg 10^4\,\GeV$.
If $g_s$ is too large, unphysical negative values
of~(\ref{Eq:AxionTPRate}) are encountered for $E\lesssim T$ that lead
to an artificial suppression of \eqref{Eq:AxionYieldTP} as indicated
by the logarithmic factor; cf.~Sec.~3
in~Ref.~\cite{Brandenburg:2004du} where a similar behavior is
discussed in more detail. Nevertheless, we think that the applied
methods still allow presently for the most reliable treatment since
they respect gauge invariance.

Note that~(\ref{Eq:AxionYieldTP}) is only valid when axion
disappearance processes can be neglected. In scenarios in which $\TR$
exceeds $\TD$, this is not justified since there has been an early
period in which axions were in thermal equilibrium. In this period,
their production and annihilation proceeded at equal rates.
Thereafter, they decoupled as hot thermal relics at $\TD$, where all
standard model particles are effectively massless. The present yield
of those thermal relic axions is then given by
$\Yaxeq = \naxeq/s \approx 2.6 \times 10^{-3}$.
In Fig.~\ref{Fig:AxionYield} this value is indicated by the horizontal
lines. In fact, the thermally produced yield cannot exceed $\Yaxeq$.
In scenarios with $\TR$ such that~(\ref{Eq:AxionYieldTP}) turns out to
be close to or greater than $\Yaxeq$, disappearance processes have to
be taken into account. The resulting axion yield from thermal
processes will then respect $\Yaxeq$ as the upper limit.
For example, for $\fax=10^9\,\GeV$, this yield would show a dependence
on the reheating temperature $\TR$ that is very similar to the one
shown by the dotted line in Fig.~\ref{Fig:AxionYield}. The only
difference will be a smooth transition instead of the kink at
$\YaxTP=\Yaxeq$.

\section{Axion decoupling temperature}

The kinks in
Fig.~\ref{Fig:AxionYield} indicate the critical $\TR$ value which
separates scenarios with thermal relic axions from those in which
axions have never been in thermal equilibrium. Thus, for a given
$\fax$, this critical $\TR$ value allows us to extract an estimate of
the axion decoupling temperature $\TD$. 
We find that our numerical results are well described by
\be
\TD
\approx
9.6\times 10^6\,\GeV\left(\frac{\fax}{10^{10}\,\GeV}\right)^{2.246}
\ .
\label{Eq:TDFit}
\ee
In a previous study~\cite{Masso:2002np}, the decoupling of axions that
were in thermal equilibrium in the QGP was calculated. 
When following~\cite{Masso:2002np} but including~(\ref{Eq:gS_SM}),
we find that the temperature at which the axion yield from thermal
processes started to differ by more than 5\% from $\Yaxeq$ agrees
basically with~(\ref{Eq:TDFit}).
The axion interaction rate~$\Gamma$ equals $H$ already at temperatures
about a factor 4 below~(\ref{Eq:TDFit}) which, however, 
amounts to a different definition of $\TD$.

% __________________________________________________________________
\begin{figure}[b]
\centering
\includegraphics[width=0.48\textwidth,clip]{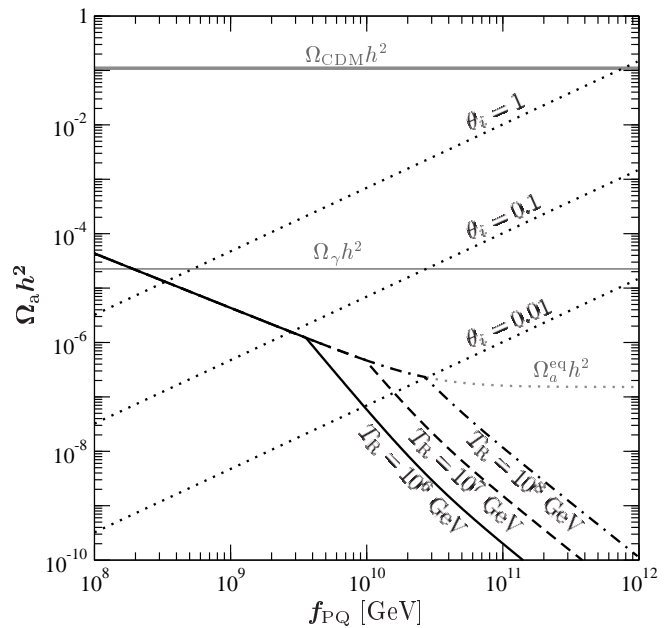}
\caption{The axion density parameter from thermal processes for
  $\TR=10^6\,\GeV$ (solid), $10^7\,\GeV$ (dashed) and $10^8\,\GeV$
  (dash-dotted) and the one from the misalignment mechanism for
  $\theta_i=1$, $0.1$, and $0.01$ (dotted). The density parameters for
  thermal relic axions, photons, and cold dark matter are indicated,
  respectively, by the gray dotted line ($\Omegaaxeq h^2$), the gray
  thin line ($\Omega_\gamma h^2$), and the gray horizontal bar
  ($\Omega_{\mathrm{CDM}} h^2$).}
\label{Fig:AxionOmega}
\end{figure}
% __________________________________________________________________
%

\section{Axion density parameter}

Since thermally produced axions have basically 
a thermal spectrum also, we find that the density parameter
from thermal processes in the primordial plasma can be described
approximately by
\begin{equation}
  \Omega_{\axion}^{\mathrm{TP/eq}} h^2 
  \simeq
  \sqrt{\langle p_{\axion,0} \rangle^2 + \maxion^2}\,
  Y_{\axion}^{\mathrm{TP/eq}}\,
  s(T_0) h^2/\rho_c
\label{Eq:OmegaTP}
\end{equation}
with the present average momentum
$\langle p_{\axion,0}\rangle=2.701\,T_{\axion,0}$ 
given by the present axion temperature 
$T_{\axion,0}=0.332\,T_0\simeq 0.08~\meV$, 
where $T_0\simeq 0.235~\meV$ is the present cosmic microwave
background temperature, $h\simeq 0.7$ is Hubble's constant in units of
$100~\mathrm{km}/\mathrm{Mpc}/\mathrm{s}$ and
$\rho_c/[s(T_0)h^2]=3.6~\eV$.
A comparison of $T_{\axion,0}$ with the axion mass
$\maxion\simeq 0.6~\meV\,(10^{10}\,\GeV/\fax)$
shows that this axion population is still relativistic today for
$\fax\gtrsim 10^{11}\,\GeV$.

In Fig.~\ref{Fig:AxionOmega} the solid, dashed, and dash-dotted lines
show
$\Omega_{\axion}^{\mathrm{TP/eq}}h^2$
for $\TR=10^6$, $10^7$, and $10^8\,\GeV$, respectively. In the $\fax$
region to the right (left) of the respective kink, in which $\TR<\TD$
($\TR>\TD$) holds,
$\Omega_{\axion}^{\mathrm{TP\,(eq)}}h^2$
applies, which behaves as 
$\propto\fax^{-3}$ ($\fax^{-1}$) for $\maxion\gg T_{\axion,0}$
and as 
$\propto\fax^{-2}$ ($\fax^{0}$) for $\maxion\ll T_{\axion,0}$.
The gray dotted curve shows $\Omega_{\axion}^{\mathrm{eq}} h^2$ for
higher $\TR$ with $\TR>\TD$ and also indicates an upper limit on the
thermally produced axion density.
Observation of those axions will be extremely challenging.
Even $\Omega_{\axion}^{\mathrm{eq}} h^2$ stays well below the cold
dark matter density $\Omega_{\mathrm{CDM}} h^2\simeq 0.1$ (gray
horizontal bar) and also below the photon density $\Omega_\gamma
h^2\simeq 2.5\times 10^{-5}$ (gray thin line)~\cite{Amsler:2008zzb} in
the allowed $\fax$ range~(\ref{Eq:fa_Limit}). There, the current
hot dark matter limits are also safely 
respected~\cite{Hannestad:2010yi}.

In cosmological settings with $\TR>\TD$, axions produced
nonthermally before axion decoupling (e.g., in inflaton decays) 
will also be thermalized resulting in $\Omegaaxeq h^2$.
The axion condensate from the misalignment mechanism, however, is not
affected---independent of the hierarchy between $\TR$ and
$\TD$---since thermal axion production in the QGP is negligible at
$T\lesssim 1~\GeV$.
Thus, the associated density 
$\Omega_a^{\mathrm{MIS}}h^2\sim
0.15\,\theta_i^2(\fax/10^{12}\,\GeV)^{7/6}$~\cite{Sikivie:2006ni,Kim:2008hd,Beltran:2006sq}
can coexist with $\Omega_{\axion}^{\mathrm{TP/eq}}h^2$ and is governed
by the misalignment angle~$\theta_i$ as illustrated by the dotted
lines in Fig.~\ref{Fig:AxionOmega}.
Thereby, the combination of the axion cold dark matter condensate with
the axions from thermal processes,
$\Omega_{\axion}h^2=\Omega_a^{\mathrm{MIS}}h^2+\Omega_{\axion}^{\mathrm{TP/eq}}h^2$,
gives the analog of a Lee--Weinberg curve.
Taking into account the relation between $\fax$ and $\maxion$, this is
exactly the type of curve that can be inferred from
Fig.~\ref{Fig:AxionOmega}.
Here our calculation of thermal axion production 
in the quark-gluon plasma allows us
to cover, for the first time, 
also cosmological settings with $\TR<\TD$.

\medskip

\begin{acknowledgments}
  We are grateful to Thomas Hahn, Josef Pradler, Georg Raffelt, and
  Javier Redondo for valuable discussions.
  This research was partially supported by the Cluster of Excellence
  ``Origin and Structure of the Universe.''
\end{acknowledgments}
%
% __________________________________________________________________
%

%
% __________________________________________________________________
%
\end{document}